\documentclass{article}
\usepackage{times}
\usepackage{graphicx} 
\usepackage{subfigure} 
\usepackage{microtype}
\usepackage[sort&compress]{natbib}
\usepackage{algorithm}
\usepackage{algorithmic}
\usepackage{amsmath}
\usepackage{hyperref} 
\usepackage{flushend}
\usepackage{booktabs}
\usepackage{multirow}
\usepackage{booktabs}

\let\oldtheequation\theequation
\makeatletter
\def\tagform@#1{\maketag@@@{\ignorespaces#1\unskip\@@italiccorr}}
\renewcommand{\theequation}{(\oldtheequation)}
\makeatother 

\usepackage[accepted]{icml2014ed}

\icmltitlerunning{}

\begin{document} 
\setcitestyle{square}
\onecolumn
\icmltitle{Protein Secondary Structure Prediction with Long Short Term Memory Networks}

\icmlauthor{S\o ren Kaae S\o nderby$^\text{1}$}{soren.sonderby@bio.ku.dk}
\icmlauthor{Ole Winther$^\text{1,2}$}{olwi@dtu.dk}
\icmladdress{$^\text{1}$ Bioinformatics Centre, Department of Biology, University of Copenhagen, Copenhagen, Denmark\\ $^\text{2}$ Department for Applied Mathematics and Computer Science, Technical University of Denmark (DTU), 2800 Lyngby, Denmark}

\icmlkeywords{Protein Secondary structure prediction, LSTM, RNN}

\vskip 0.3in

\begin{abstract}
Prediction of protein secondary structure from the amino acid sequence is a classical bioinformatics problem. Common methods use feed forward neural networks or SVM's combined with a sliding window, as these models does not naturally handle sequential data. Recurrent neural networks are an generalization of the feed forward neural network that naturally handle sequential data. We use a bidirectional recurrent neural network with long short term memory cells for prediction of secondary structure and evaluate using the CB513 dataset. On the secondary structure 8-class problem we report better performance (0.674) than state of the art (0.664). Our model includes feed forward networks between the long short term memory cells, a path that can be further explored.
\end{abstract}

%\tableofcontents
%!TEX root = report.tex
\section{INTRODUCTION} % (fold) 
Recently Long Short Term Memory (LSTM) \citep{Hochreiter1997} recurrent neural networks (RNN) have shown good performance in a number of tasks, including machine translation \citep{Sutskever2014}, and speech recognition \citep{Graves2014}. This paper uses the LSTM for prediction of protein secondary structure. Many machine learning algorithms have been applied to this problem: \citealt{Qian1988} introduced neural networks, \citealt{Jones1999} discovered that the use of evolutionary information, through position specific scorring matrices, improved performance, and \citealt{Baldi1999} introduced RNN's for secondary structure prediction. Recent work includes conditional random fields hybrid models \citep{Maaten2011,Peng2009,Wang2011} and generative stochastic networks \citep{Troyanskaya2014}.\\\\%
A common approach to secondary structure prediction is to use a non-sequential model, typically feed-forward neural networks or SVM's \citep{Jones1999,Hua2001}. These models are not ideal for classifying data which cannot naturally be presented as a vector of fixed dimensionality, why a sliding window approach is typically used to circumvent this problem. Window based models can only learn dependencies within the input window, recent methods for learning other dependencies includes conditional random field hybrid models. RNN's can be applied to sequential data of any length, and should theoretically be able to learn longterm dependencies. In practice RNN's suffer from exploding or vanishing gradients \citep{Bengio1994}, and on a secondary structure prediction task \citealt{Baldi1999} reported that their RNN's were only able to learn dependencies of  $\pm15$ amino acids relative to the target. The LSTM cell was invented to solve the vanishing gradients problem and enables the network to learn dependenceis over 100's of time steps. The contribution of this paper is the application of bidirectional LSTM networks \citep{Graves2012} to protein secondary structure prediction. Our model architecture uses feed-forward neural networks for concatenation of predictions from the forward and backward networks in the bidirectional model and the model also includes feed-forward neural networks between hidden states in the recurrent network, See figure \ref{fig:lstm_model}. The use of feed-forward neural networks "inside" the reucrrent neural network has also been explored by \citep{Pascanu2013a}. This work primarily differs from the work by \citealt{Baldi1999} in the introduction of the LSTM cell, the availability of much larger datasets and the possibility of training larger models by using a GPU.

%!TEX root = report.tex
\section{MATERIALS AND METHODS}%
\subsection{Model}
\begin{figure}[tb]
	\begin{center}
		\includegraphics[width=0.7\textwidth]{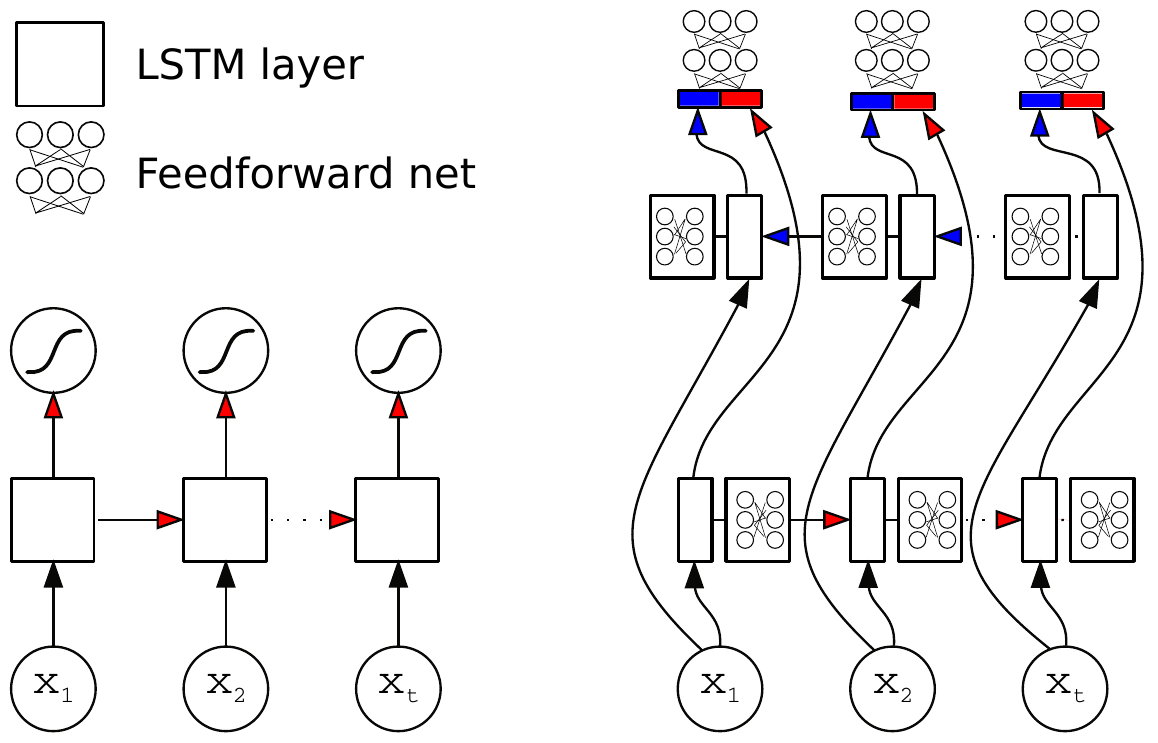}
	\end{center}
	\caption{Unrolled recurrent neural networks. \textit{left}: Unidirectional LSTM with a single layer. \textit{right}: Bidirectional LSTM with single layer. The forward LSTM (red arrows) starts at time $1$ and the backwards LSTM (blue arrows) starts at time $n$, then they go forwards and backwards respectively. The errors from the forward and backward nets are combined using a feed forward net and the result is used for back propagation. Note the feedforward nets between time slices.
	The figure shows a single layer model, but the model is easily extended with more layers. Adapted from \citep{Graves2012}.}
	\label{fig:lstm_model}
\end{figure}

\begin{figure}[tb]
	\begin{center}
		\includegraphics[width=0.5\textwidth]{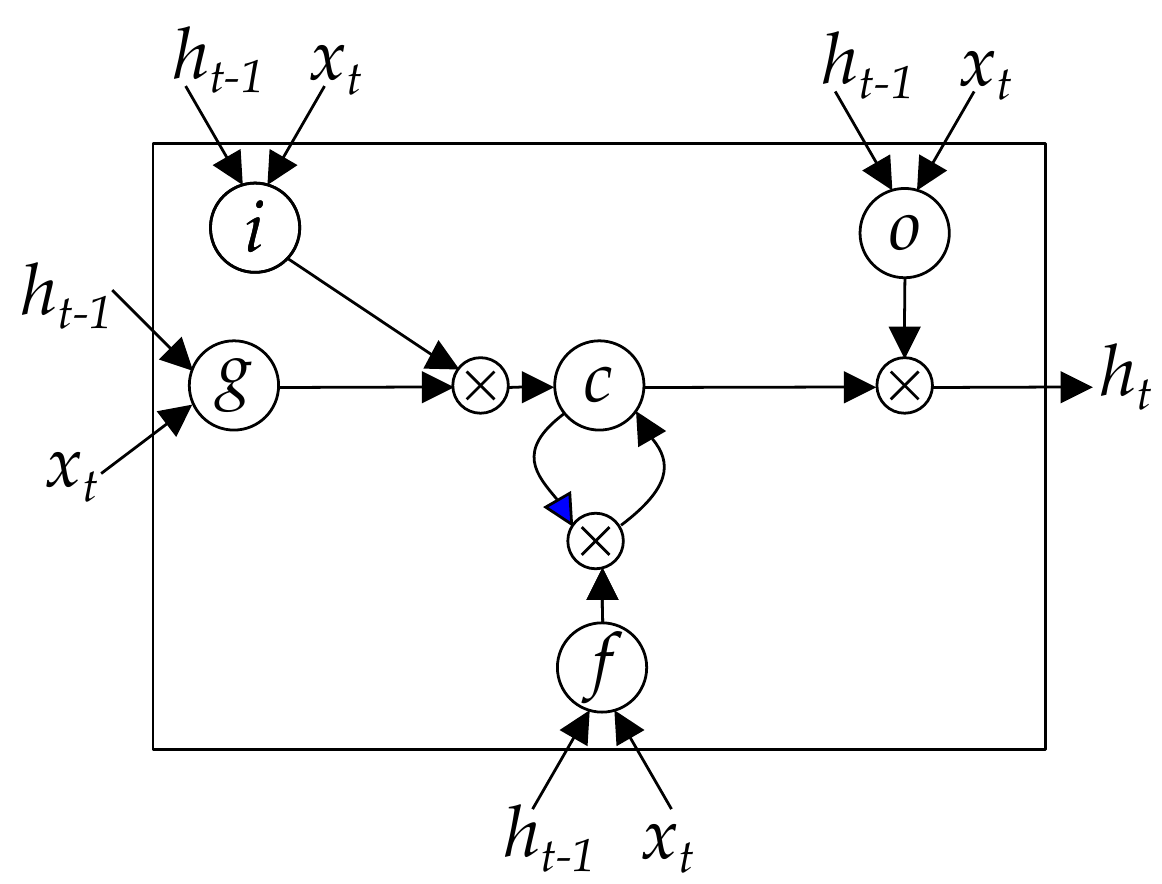}
	\end{center} 
	\caption{LSTM memory cell. \textit{i}: input gate, 
	         \textit{f}: forget gate, 
	         \textit{o}: output gate, 
	         \textit{g}: input modulation gate, 
	         \textit{c}: memory cell. Blue arrow heads are $c_{t-1}$ and red arrow heads are $c_t$.
	         The notation corresponds to equations \ref{eq:it} to \ref{eq:xt} such that $W_{xo}$ is wights for $x$ to output gate
	         and $W_{hf}$ is weigts for $h_{t-1}$ to forget gates etc. Adapted from \citep{Zaremba2014a}.}
	\label{fig:lstm_cell}
\end{figure}

The LSTM cell is implemented as described in \citep{Graves2013}, however without peepholes, because recent papers have shown good performance without peepholes \citep{Zaremba2014a,Zaremba2014b,Sutskever2014}. When predicting target $x_t$ a (forwards) RNN only know the past sequence, $x_1 ... x_t$. In tasks where the entire sequence is known beforehand, e.g. secondary structure prediction, this is not desirable. \citealt{Schuster1997} introduced the bidirectional RNN as an elegant solution to this problem. One trains two separate RNN's, the forward RNN starts the recursion from $x_1$ and goes forwards, the backwards model starts at $x_n$ and goes backwards. The predictions from the forward and backward networks are combined and normalized, see Figure \ref{fig:lstm_model}. The standard method for combining the forward and backward models is to normalize the activations from each layer in a softmax layer \citep{Graves2012}. We expand the standard stacked bidirectional LSTM model by introducing a feed-forward network responsible for concatenating the output from the forward and backward networks into a single softmax prediction. Secondly we expand the model by inserting a feed-forward network between recurrent hidden states, see equation \ref{eq:ff}, along with shortcut connections between the recurrent hidden layers. Similar ideas have been explored for RNN's by \citep{Pascanu2013a}. Figure \ref{fig:lstm_cell} shows a LSTM cell. Equation \ref{eq:it} to equation \ref{eq:xt} describes the forward recursions for a single LSTM layer, $h_{t-rec}$ is forwarded to the next time slice and $h_t$ is passed upwards in a multilayer LSTM. % The LSTM cells are regularized with dropout on non-recurrent connections as suggested in \citealt{Zaremba2014}.
\begin{align}
i_t  &= \sigma(x_{t}W_{xi} + h_{t-1}W_{hi}  + b_{i})  \label{eq:it}\\	
f_t  &= \sigma(x_{t}W_{xf} + h_{t-1}W_{hf}  + b_{f}) \\
o_t  &= \sigma(x_{t}W_{xo} + h_{t-1}W_{ho}  + b_{o}) \\
g_t  &= \tanh(x_{t}W_{xg} + h_{t-1}W_{hg} + b_{g}) \\
c_t  &= f_t \odot c_{t-1} +  i_t \odot g_t  \\
h_t  &= o_t \odot \tanh(c_{t}) \\
h_{t-rec} &= h_t + feedforwardnet(h_t)  \label{eq:ff}\\
\sigma(z)    &= \frac{1}{1+\exp(-z)} \\
%D    &= \text{Elementwise set to zero with probability } p \\
\odot &: \text{Elementwise multiplication} \\
x_t  &: \text{input from the previous layer: } h_t^{l-1}  \label{eq:xt}
\end{align}
%
%
%\begin{align}
%r_t  &= \sigma(D(x_{t}W_{xr}) + h_{t-1}W_{hr}  + b_{r})  \label{eq:ir}\\	
%z_t  &= \sigma(D(x_{t}W_{xz}) + h_{t-1}W_{hz}  + b_{z}) \\
%h_c  &= \tanh(D(x_{t}W_{xh}) + (r_t \odot h_{t-1})W_{hh} + b_h) \\
%h_t  &= (1-z_t)\odot h_{t-1} + z_t \odot h_c   \label{eq:ht} 
%\end{align}
%
%
\subsection{Data}
We use the dataset from \citealt{Troyanskaya2014}\footnote{\url{http://www.princeton.edu/~jzthree/datasets/ICML2014}}. The dataset consists of amino acid sequences labeled with secondary structure. Sequences and structures were downloaded from PDB and annotated with the DSSP program \citep{Kabsch1983}. In the literature it is common to map the 8-class DSSP output (Q8) to helix, sheets and coils (Q3), see Table \ref{tab:struclabels}. We use the original 8-class output, which is a harder problem. Each amino acid is encoded as an 42 dimensional vector, 21 dimensions for orthogonal encoding and 21 dimensions for sequence profiles. For further descriptions see \citealt{Troyanskaya2014}. The full dataset has 6128 non-homologous sequences (identity less than 30\%). This set is further filtered such that no sequences has more than 25\% identity with the CB513 dataset \citep{Cuff1999}. The dataset is divided into a training (n=5278) and a validation set (n=256), the CB513 dataset is used for testing.% 
\begin{table}[tb]
	\caption{Description of protein secondary structure classes and class frequencies in the dataset. In the litterature the 8-class DSSP output is typically mapped to 3 classes. The 8 to 3 class mappings are included for reference.}
	\label{tab:struclabels}
	\begin{center}
		\begin{tabular}{lllll}
		\hline
		\textbf{8-class (Q8)}&\textbf{3 class (Q3)} &\textbf{Frequency}& \textbf{Name} \\
		\hline
			 H&  H & 0.34535 & $\alpha$-helix\\
			 E&  E & 0.21781 & $\beta$-strand\\
			 L&  C & 0.19185 & loop or irregular\\
			 T&  C & 0.11284 & $\beta$-turn\\
			 S&  C & 0.08258 & bend\\
			 G&  H & 0.03911 & $3_{10}$-helix\\
			 B&  E & 0.01029 & $\beta$-bridge\\
			 I&  C & 0.00018 & $\pi$-helix\\
		\hline
		\end{tabular}
	\end{center}
\end{table}%
\subsection{Experimental setup}
 The LSTM is implemented in Theano \citep{Bastien2012} using the Lasagne library\footnote{\url{https://github.com/benanne/Lasagne}}. 
 %Dropout is applied to the non-recurrent connections as suggested in \citealt{Zaremba2014}. 
The model has 3 layers with either 300 or 500 LSTM units in each layer. 
 %50\% dropout is applied to the non-recurrent connections. 
The feed-forward network, eq. \ref{eq:ff}, is a two layer ReLU network with 300 or 500 units in each layer, this network has skip connections. The output from the bidirectional forwards and backwards networks are concatenated into a single vector which is passed through a two layer ReLU network with 200 or 400 hidden units in each layer. The concatenation network is regularized using 50\% dropout. In the LSTM cells all initial weights are sampled uniformly between -0.05 and 0.05 and biases are initialized at zero. In the fully connected layers weights are initialized using Lasagne's default settings. The LSTM initial hidden and cell states are learned. The learning rate is controlled with AdaDelta using default settings ($\rho =0.95$, $\epsilon=10^{-6}$)\citep{Zeiler2012}. After each epoch we calculate the norm of the gradients updates divided by the batch size:

\[ norm_2 = \begin{Vmatrix} \frac{\text{gradient updates}}{\text{batch size}}\end{Vmatrix}_2 \]
If the norm exceeds 0.5 all gradients are scaled with $\frac{0.5}{norm_2}$. The batch size is 128. 
%!TEX root = report.tex
\section{RESULTS} % (fold)
The LSTM network has a correct classification rate of 0.674, better than current state of the art performance achieved by a generative stochastic network (GSN) \citep{Bengio2013a,Troyanskaya2014} and a conditional neural field (CNF) \citep{Peng2009,Lafferty2001}. Furthermore the LSTM network performs significantly better than the bidirectional RNN (BRNN) used in SSpro8 having a correct classification rate of 0.511 \citep{Pollastri2002}, see Table \ref{tab:results}. 
\begin{table}[tb]
	\caption{Test set per amino acid accuracy for CB513. $^*$Reported by \citealt{Wang2011}}
	\label{tab:results}
	\begin{center}
		\begin{tabular}{ll}
		\hline
		 & \textbf{Q8 accuracy}\\
		\hline
			 \citep{Pollastri2002} (BRNN)$^*$             					&$0.511$ \\                 
			 \citealt{Wang2011} (CNF - 5-model ensemble)      &$0.649$ \\
			 \citealt{Troyanskaya2014} (GSN)					&$0.664$ \\
			 LSTM small									&$0.671$\\
			 LSTM large                                 &$\mathbf{0.674}$ \\

		\hline
		\end{tabular}
	\end{center}
\end{table}
%!TEX root = report.tex
\section{DISCUSSION AND CONCLUSION}%
We used the LSTM RNN for prediction of protein secondary structure. To our knowlegde the 
CB513 performance of 0.674 is currently state-of-the-art. Comparision with the SSpro8 method shows that the LSTM significantly improves the performance. Similarly the LSTM performs bettter than both Conditional neural fields and GSN methods. Inspired by \citealt{Pascanu2013a}
we used a feedforward network between the recurrent connections. We showed that a LSTM with 
this architecture and a feedforward neural net for concatenation of the forward and 
backward nets performs significantly better than existing methods for secondary structure prediction. Future work includes investigation of different architectures for the feedforwards networks. 

%!TEX root = report.tex
\section{AUTHORS CONTRIBUTIONS}
SS is PhD student under the supervision of OW. SS developed the model and performed the
experiments. Both authors read and approved the final version of the article.

\section{ACKNOWLEDGEMENTS}
We gratefully acknowledge the support of NVIDIA Corporation with the donation of the Tesla K40 GPU used for this research. We wish to acknowledge funding from the Novo Nordisk Foundation.

\begin{footnotesize}
\bibliographystyle{icml2014}
\bibliography{library.bib}
\end{footnotesize}

\end{document}